\documentclass[times,twocolumn,final,authoryear]{elsarticle}

\usepackage{prletters}
\usepackage{framed,multirow}
\usepackage{threeparttable}

\usepackage{amssymb}
\usepackage{latexsym}
\usepackage{amsmath}
\usepackage{url}
\usepackage{xcolor}
\definecolor{newcolor}{rgb}{.8,.349,.1}

\usepackage{hyperref}
\usepackage{pdfpages}
\usepackage{booktabs} 
\usepackage{longtable}
\usepackage{algorithmic}
\usepackage{multirow}
\usepackage[T1]{fontenc}

\usepackage[ruled]{algorithm2e} 

\SetAlFnt{\small}
\SetAlCapFnt{\small}
\SetAlCapNameFnt{\small}
\SetAlCapHSkip{0pt}
\IncMargin{-\parindent}

\newcommand{\argmin}{\arg\!\min}

\journal{Pattern Recognition Letters}

\begin{document}

\thispagestyle{empty}

\clearpage

\ifpreprint
  \setcounter{page}{1}
\else
  \setcounter{page}{1}
\fi

\begin{frontmatter}

\title{Software Expert Discovery via Knowledge Domain Embeddings in a Collaborative Network}

\author[1]{Chaoran \snm{Huang}\corref{cor1}} 
\cortext[cor1]{Corresponding author: 
  Tel.: +61-2-938-56909;}
\ead{chaoran.huang@unsw.edu.au}
\author[1]{Lina \snm{Yao}}
\author[2]{Xianzhi \snm{Wang}}
\author[1]{Boualem \snm{Benatallah}}
\author[1]{Xiang \snm{Zhang}}
\address[1]{UNSW Sydney, NSW 2052, Australia}
\address[2]{Singapore Management University, Singapore 178902}
\received{1 May 2013}
\finalform{10 May 2013}
\accepted{13 May 2013}
\availableonline{15 May 2013}
\communicated{C.Huang}

\begin{abstract}
    Community Question Answering (CQA) websites can be claimed as the most major venues for knowledge sharing, and the most effective way of exchanging knowledge at present. 
    Considering that massive amount of users are participating online and generating huge amount data, management of knowledge here systematically can be challenging.
    Expert recommendation is one of the major challenges, as it highlights users in CQA with potential expertise, which may help match unresolved questions with existing high quality answers while at the same time may help external services like human resource systems as another reference to evaluate their candidates.
    In this paper, we in this work we propose to exploring experts in CQA websites.
    We take advantage of recent distributed word representation technology to help summarize text chunks, and in a semantic view exploiting the relationships between natural language phrases to extract latent knowledge domains. By domains, the users' expertise is determined on their historical performance, and a rank can be compute to given recommendation accordingly. In particular, Stack Overflow is chosen as our dataset to test and evaluate our work, where inclusive experiment shows our competence.
\end{abstract}


\end{frontmatter}


\section{Introduction}
\label{sec:introduction}
Community Question and Answering (Q\&A) websites is one of the most common ways of online collaboration, which may be the most effective knowledge sharing approach. 
Those websites are designed to depend on users' participations. 
Typically in CQA sites, a requester can post a problem, waiting for contributors to post solutions, while at the same time, other users can browser, comment and vote for a best answer.
Basically, the ``wisdom of crowds'' do help to solve  strenuous problems, yet such a diagram is meant to lose control of quality of posts, not to mention the participation of users itself.

Let's take one successful CQA website with more than 5 million users, Stack Overflow\footnote{\url{http://stackoverflow.com}}\footnote{Data used in this work are from ``Stack Overflow public dump'' at \\ \url{http://archive.org/download/stackexchange}} as an example.
Here, one( the requester) can ask questions and others with specific skills may answer voluntarily; He/she may also add tags to help answerers to locate answers of interest, and the viewers may vote up or down to questions and answers; Additionally the requester can nominate one answer as the accepted answer which satisfies him most. The system can be claimed relatively productive in exchanging domain knowledge.

Accessed on 10 March 2016, where 6,120,191/11,053,469 questions have accepted answers. And there are approximately 27\% of the 11 million questions have no activity at all; where among the rest, nearly half have no accepted answers. 
As studied in \cite{wang2018survey}, expert recommendation can potentially help to boost user contribution, by recommending specialists to those untouched or unresolved problems, which in another hand also secures post quality since unrelated users or non-professional are compared less likely to be pushed to give an answer. And given data available at this scale, it is particularly attainable to given recommendations based on historical posts.

\begin{figure*}
  \centering
  \label{fig:badTags}
  \includegraphics[width=0.95\linewidth]{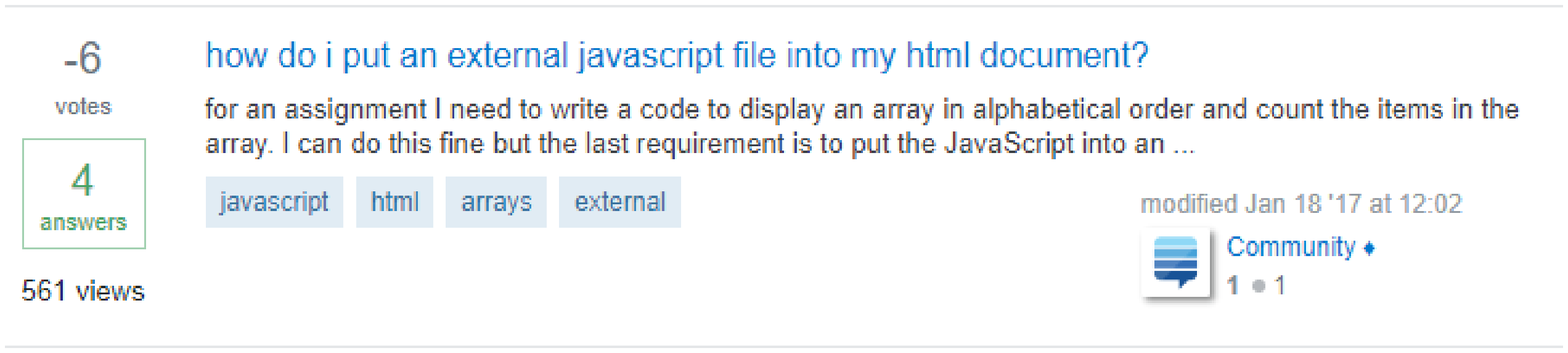}
  \caption{An example of Inappropriate tags assigned by user.}
\end{figure*}

The above 
example reveal the value and potential applications of expert recommendation in online CQA website, 
with an objective 
that is to determine specific domain the question posts lay on. Classifying posts by its knowledge domain in one hand helps others to quick direct in existing posts and find post of interest; in another hand, it helps to present new question to be answered by the right people, that is, those who are browsing post of the same category and those who sharing same profession.

As mentioned before, like many CQA systems employing tag system to help, Stack Overflow is no exception. Question raisers can at most assign 5 tags to a post. To our intuition and observation, question raisers are often not quite familiar with the domain where he/she ask the question, yet the tags are assigned or even new tags can be created by them, which can be inaccurate or misleading(An example can be Fig~\ref{fig:badTags}).  
Inappropriate tags can lead to a post being unnoticed or less attractive, previous study as \cite{guo2008tapping} shows this would cause the post not resolved. Although it is arguable that some existing works, \cite{guo2008tapping, dong2015predicting}, are based on those tags, we propose to exact the latent domains to avoid the prospective inadequacies. More often, existing works like \cite{chiang2012exploring,zhang2007expertise,Hanrahan:2012:MPD:2141512.2141550,Riahi:2012:FEU:2187980.2188202} build profiles for each user, and in our case, this is not feasible due to our monumental data scale.

Hence, here we propose a framework for recommending experts in collaborative networks, which relies on knowledge domain embeddings produced from user generated content. Given a query consists of either one or more keywords or phrases, the out put will be a ranked list of expert related to the query. 
We first prepare and train language embeddings on the CQA text data, which are sent to be clustered as derived domains of knowledge. High quality posts are therefore picked and assigned correspondingly to domains, where experts, i.e. authors of those posts, can be inferred given a query. As a extended journal version of previous work(\cite{8029777}), 
following contributions can be claimed:  

\begin{itemize}
  \item We take advantage of recent distributed word representation technology to help summarize text chunks; The embeddings are utilized both in semantically and numerically;
  \item We explore the relationships between natural language phrases in a semantic view, to extract latent knowledge domains, where the chosen of domain is analysis and assessed systematically;
  \item Users' expertise is determined on their historical performance, while the potential data sparseness issue is alleviated by matrix factorization approach;
  \item Our method is test and evaluated with a relative large scale dataset with comprehensive experiments, where preferable output is generated.
\end{itemize}

The remainder of the article is structured as follows. Section~\ref{sec:relatedworks} briefs related existing works; Section~\ref{sec:methodology} introduce and explains our framework in details; Section~\ref{sec:experiment} describes experiment set-up and procedure, along with analysis of results and evaluation; Finally Section~\ref{sec:conclusion} concludes this work with remarks.

\section{Related Works}
\label{sec:relatedworks}
Expert recommendation is always a long-standing and important research topic of  information retrieval and knowledge management. 
And the popularity of online communities accelerates the trend.
Largely due to the limit in computing power and the absence of study in neural networks, earlier works, like ~\cite{jurczyk2007hits,wang2002ranking}, are often based on conventional recommender systems and focus on user link analysis and user behaviors, which based on the assumption that experts are likely to have links and interactive with other experts. Instead shed light solely on users, recent works are more complex and multitudinous.

\cite{chiang2012exploring} propose to recommend by a graph-based model.
They rely on the user browsing logs and claim language dependence can cause problems in graph based recommender models in Q\&A systems and make user generated contents not reliable.
Also, they identify users browse not only Q\&A pages in a website. The Continuous-time Markov model is applied to generate a so called ``QA Latent Browsing Graph", which can help to alleviate data sparsity issue, and based which, ``Latent Browsing Rank" and ``Latent Browsing Rank Recommendation" are proposed as the importance score and recommendation module. They hence can make recommendations accordingly.

Apart from computing user expertise, question difficulty can also be a reference to infer experts.
Both are interesting to Hanrahan and his group. 
In \cite{Hanrahan:2012:MPD:2141512.2141550}, their research propose to reveal question difficulty by mining question-answer events, and which is combined with the reputation score Stack Overflow provides. Alternatively, user events as giving up-votes and down-votes can also be utilized to determine user expertise(\cite{zhang2007expertise}).
\cite{Riahi:2012:FEU:2187980.2188202} build user profiles to rank users. They reveal underlying connections between users and questions and appropriate users are recommended based on their interests. Models like Latent Dirichlet Allocation (LDA) and Segmented Topic Model (STM) for clustering are also inclusively compared during their experiments.

User performance is usually the only element considered in past works for picking answerers to questions. Yet, better results can be achieved in our perspective, where the textual is also our concern, as it contains substantial relevance information and recent language processing technology enabled this. 
\cite{dong2015predicting} classify users according to the similarities between their topics and questions for a better recommendation. However, user tags and uses user authority are selected as metrics, where tags in some cases may not be trust. \cite{guo2008tapping} also come up with a  topic-based method, and their general idea is to either investigate the answers to questions that are similar to unsolved one, or study user's history to determine his/her expertise.

\begin{figure*}[!ht]
  \centering
  \includegraphics[width=0.95\linewidth]{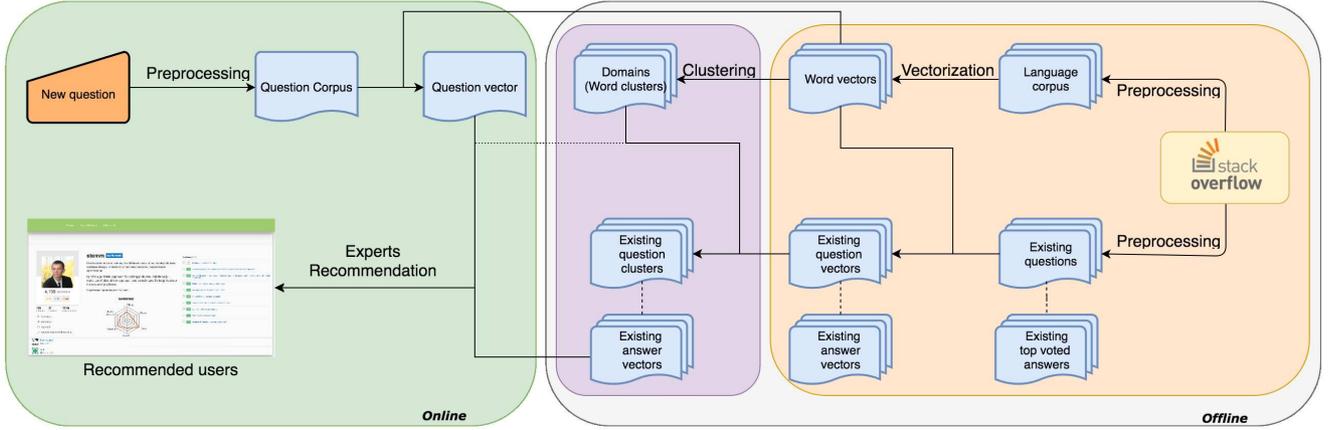}
  \caption{Framework of Answerer Recommendation}
  \label{fig:framework}
\end{figure*}

\section{Our Approach}
\label{sec:methodology}
\subsection{Problem Formulation}
Above two examples in introduction shed light on the value and application of expert recommendation system, and we can identify that they share common objectives, which enable us to recommend experts: 
\begin{enumerate}
  \item determine the specific knowledge domain of problem, which can help to assign contributor accordingly or help to know the users active domain; 
  \item evaluate the expertise in the domains where the user have involvements.
\end{enumerate}

In this section, we discuss how to achieved the above objectives.

Let $U=\{u_1, u_2,...,u_m\}$ be the set of users in our dataset, $E =\{e_1, e_2,...,e_n\} \subset U$ is a subset of $U$ denotes expert users. For posts $P$ in the dataset, each post have an author $\alpha \in U$ accordingly, and the post can be either a question $q_i \in Q \subset P={q_1, q_2,...,q_r}$ or an answer $a_i \in A \subset P$, where $A$ is the set of answers with authors $\mathcal{A}$, with a score $\sigma$. 
A post $p_i \in P$ can have a domain topic $t \in T=\{t_1, t_2,...t_s\}$. 
Given a query $q$, we can claim the topic of query $q$ is $t_q$, and for which, a most similar existing question $q\prime$ can be identified, with a domain topic $t_{q\prime}$. 
Considering that we have a huge number of questions($r$) available with a limited number of domain topics($s$, where $r>>s$), we can safely assume $t_q = t_{q\prime}$. 
Our intuition here is that if existing question $q\prime$ is satisfied by its author, and the top-$k$ most high score answers($A\prime=\{a_u, a_{u+1},..., a_{u+k-1}\}$) of question $p\prime$ all have a decent score($min({\sigma}_u, {\sigma}_{u+1},..., {\sigma}_{u+1-k}) > v$, where $v$ is the threshold controls the answer quality), we claim potential experts $E\prime$ is among the authors of those answers, that is, $E\prime=\{{\alpha}_u, {\alpha}_{u+1}, {\alpha}_{u+k-1}\}$.

Figure~\ref{fig:framework} illustrates the framework of our approach, and in summary, the method consists of three major stages:
\begin{enumerate}
  \item {\bf Post Representation}
    In the first stage the task is to formulate our dataset to be ready for computing. Our raw data is plain text in natural language, and we propose to represent the data in vectors, which can be more friendly for later procedures.     
  \item {\bf Expert Domain Exaction} 
    As aforementioned, we claim that user-generated tags are not reliable, and can have negative influence in some cases. In this stage, we propose to extract domains automatically by employing clustering techniques upon the post representations.
  \item {\bf Expert Recommendation} 
    In the last stage we produce the recommendation. We firstly using the same procedure in the above two procedures to determine the domain of the given query, and assign a related existing question in the domain to the query. A list of potential experts to the query can be therefore inferred by the related existing question. 
\end{enumerate} 

\subsection{Post Representation}
\subsubsection{Post Preprocessing}
Bearing in mind our problem is a mining task, it is crucial to preprocessing the data first. Considering our data is website archive, removal of redundant and irrelevant information and reforming data into suitable format can be beneficial.

Symbols and annotations usually are removed in this procedure at first place, while in our case we retained some.  Note that here we are processing dump of Stack Overflow, which is a programming oriented website and where codes are quite common be included in texts. Codes themselves can be challenges to most language processors, yet, the comments in codes are usually in natural languages and can be processed. However, dependent on the programming language, various annotations can be found to comment. Here we kept tags of posts to help identify type of languages, and applying regular expressions accordingly. Additionally, we noticed that numerous html tags used to formatting posts in the texts, which also can be removed easily by regular expressions. Same procedure also applies to comments data, as well as post edit history, since given a larger number of data source may help to produce more accurate word representations.

Given our idea to recommend by post, the insurance of quality of post can be essential to secure the users inferred are confidently experts. Particularly, in this work, the assumption is satisfied questions along with their top-voted answers can be considered high quality posts, and the authors behind them are experts candidates. Such a selection will largely reduce our data  and the number of candidates safely, and bring down the massive dataset to a practical scale, while at the same time assure the necessary information at better quality. For the record, stop words are also removed during this procedure. 

\subsubsection{Word Representation}
Statically represent words or phrases using relatively lower dimensional vectors is the idea of distributional word representations. Owing to its computational complexity, despite it is invented decades ago, applications are emerging recently, and Word2Vec by \cite{mikolov2013distributed} can be credited to boost the applications of such technology. Unlike conventional method to compute distributional representations, which counting and calculate distribution of words at the whole document scale, Word2Vec go through sentences in corpus with a window, examining surroundings to learn the relationships between words closeby. In such a way, Word2Vec produce word representations by a prediction model consists of two layer neural network activated by Softmax functions. The efficiency of such model is far better while the accuracy is not compromised.

Since expertise domain is the key for inference, it is also critical to make domain extraction in the second stage flawless. In the light of our text data is almost ready, we can now remove irrelevant words. Dictionary based filter can be the simplest and fastest way, and it is also our choice here. 

\subsubsection{Post Representation}
Considering posts are selected and word representations are ready, we can move on to vectorize posts into their representations. For the two kind of posts in our study here, it is fesible to treat them with no difference in terms of representation, as documents. 

Traditional term frequency(TF, see~\cite{lee1997document}) can be one approach to modeling documents.
Specifically, given a document (post) ${d\in{D}}$ we establish a set of distinct terms ${T=\{t_1,t_2,...,t_n\}}$ that

occur in ${D}$. Then the document ${d}$ can be represented as a vector of dimension ${n}$ where each element corresponds to terms in document ${d}$ and its value is frequency of the term
denoted as ${tf(d,t)}$. Thus, we have a vector representation of document ${d}$:
\begin{equation}
  \vec{t_d}=(tf(d,t_1),tf(d,t_2),...,tf(d,t_n))
\end{equation}

TF based model assumes that the importance of a word is positively related to its occurances, i.e. frequency. This may not true in some cases, especially when the input document is short. Alternatively TF-IDF can be used to address this issue, yet disadvantage can be still claimed that by such means the semantic information are lost, and hence relationships of words are not taken into consideration. An example is that by those approaches similar words are treated with no difference than all other words.   

\begin{algorithm}[!ht]
  \caption{Our algorithm to cluster posts, it takes pre-trained word vectors ${W}$, vectorized posts ${P}$ and clustering number ${n}$ as input, and returns them in clusters}
  \label{alg:cluster}
  \begin{algorithmic}[1]
  \renewcommand{\algorithmicrequire}{\textbf{Input:}}
   \renewcommand{\algorithmicensure}{\textbf{Output:}}
   \REQUIRE ${P,n}$
   \ENSURE  ${T}$
   \\\textit{\bf Algorithm} $PostClustering(P,n)$
  
  \STATE ${T \gets Cluster(n,W)}$
  \FOR { ${t \in C}$ and ${t_{centroid} \in T_{centroid} }$}
    \STATE ${t_{centroid} \gets mean(t)}$
  \ENDFOR
  \FOR{ ${p \in P}$ }
    \FORALL{i corresponding to ${{T_i} \in T}$ \\such that minimize ${d_{cos}(p,T^i_{centroid})}$\\}
      \STATE add ${p}$ into ${T_i}$
    \ENDFOR
  \ENDFOR
  \RETURN $T$
  
  \end{algorithmic}
\end{algorithm}

We here propose to utilize the idea of TF approach, to combine it with modern word representations. Based on the pre-trained language model ${\mathcal{T}=\{\mathcal{T}_1,\mathcal{T}_2,...,\mathcal{T}_n\}}$ which represents the terms in a documents set ${D}$, given any document ${d}$ with its term frequency vector ${\vec{t_d}}$, we can summarize this document using the weighted average sum of word vectors, i.e.,
\begin{equation}
   \vec{t_d^{\prime}}=\sum{\mathcal{T}_i\times{tf(d,t_j)}}
\end{equation}
where terms ${t_j}$
correspond to word vector ${\mathcal{T}_i}$. 
Such weighting average schema have been proven effective, an example can be \cite{zhu2014weighting}.

After the vectorization of posts, the similarities among posts can be compute by distance metrics. Consine distance are employed in this research, which compare the angle of two vectors. Given two documents ${\vec{t_1}}$ and ${\vec{t_2}}$, their cosine similarity is computed by using the following equations.
\begin{equation}
  d_{cosine}= \frac{\vec{t_1}\cdot{\vec{t_2}}}{|\vec{t_1}|\times{|\vec{t_2}|}}
\end{equation}

\subsection{Expertise Domain Exaction}
Considering the massive quantity of posts we have as candidates, it would be too time consuming to calculate similarities between inputs and all the posts. Millions of times of comparison can also be argued to computational expensive. Dividing posts by their domain thus became vital. 
Traditionally domain or topics exaction are usually directly performed documents. 
\cite{beil2002frequent} proposed a term frequency based approach, which shares the same drawbacks mentioned in previous subsection. In some works, tags are used to identify domains the posts belong to.
\cite{begelman2006automated} using tagged documents as ground truth to infer the rest in a partially tagged dataset, in which they risk trust on those minor in quantity yet-may-harmful data. 
Recent studies shows semantic-based approach may have advantages in such tasks(\cite{Dumais:1998:ILA:288627.288651,Li:2008:TDC:1321784.1321938,hu2009exploiting}), especially with word vectors and document representations(\cite{kalogeratos2012text,forsati2013efficient}). 
A popular  example can be Latent dirichlet allocation(~\cite{blei2003latent}), while studies claim that LDA works not so desirable on short documents, due to its statistical natural. 

Following the idea of \cite{kalogeratos2012text}, in this work, we propose to apply clustering algorithm on the word vectors we produced before on the semantic similarities. 
Due to the vectors are representing words semantically and sentimentally, it can be testified that words in the same cluster share the same concept and thus we can use it as the domain. Since we have vectorized posts and see them as documents, it would be also meaningful and sensible to treat clusters with element words in them as documents, given only where words appear once. 
We averagely summarize words in clusters to produce centroids of clusters, which is similar to the concept of global context vectors \cite{kalogeratos2012text} proposed. Those centroids are used as representations of domains. 

\begin{algorithm}[!ht]
\caption{Our procedure to infer recommended users, it takes vectorized input question ${q}$, clustered Posts ${T}$ with its centroid set ${T_{centroid}}$ as input, and returns a list of limited top recommended users ${E}$}
\label{alg:framework}
\begin{algorithmic}[1]
\renewcommand{\algorithmicrequire}{\textbf{Input:}}
 \renewcommand{\algorithmicensure}{\textbf{Output:}}
 \REQUIRE ${q,T,T_{centroid}}$
 \ENSURE  ${E}$
 \\\textit{\bf Procedure} $UserRecommend(q,T,T_{centroid})$
\FOR{$t_{centroid} \in T_{centroid}$}
  \STATE ${d_{tq} \gets d_{cos}(q,t_{centroid}) }$
  \STATE put ${d_{tq}}$ in distance set ${D_{cq}}$
\ENDFOR
\STATE ${T_q \gets T}$ such that the corresponding ${d=\min(D_{tq})}$
\FOR{${a in C_q}$}
  \STATE compute ${d_{aq} \gets d_{cos}(a,q)}$
  \STATE put ${d_{aq}},$ into ${D_{aq}}$
\ENDFOR
\STATE initialize list $E$
\STATE ${A} \gets rank(D_{aq},\ell)$ \COMMENT{$rank(S,k)$ returns top $k$ result in set $S$}
\STATE ${E = getAuthors(A)}$ \\\COMMENT{$getAuthors(P)$ returns authors for each Posts in $P$}
\RETURN ${E}$
\end{algorithmic}
\end{algorithm}

\subsection{Expert Recommendation} 
New query for recommendation can be accepted as soon as the expert domain extraction, that is, clustering of word vectors, finished. The last stage here is aimed to output the results of expert recommendation.

Firstly, the input query goes through the same preprocessing procedure to be ready for summarizing by the weight of term frequency, into a vector in the same space of processed existing questions, as well as clusters, which represents knowledge domains. By matching the input with domains, significant number of search is skipped, while simultaneously the reduce of nonsensical computing can help to increase the chance of finding proper experts. Comparing query with posts within a cluster may still not so desirable as the number of answers can still be huge. We instead, compare the query vector to the existing questions within the cluster, and this further reduces computational resources required, and the most likely experts can still be retrieved, according to the most similar existing questions. 

The accepted answer is the one chosen by questioner. It is usually the one satisfies the questioner and the one with highest score. Users are more likely to post answers to high quality questions, which leads to our thinking on the value of unaccepted answers. We found that in rare cases, the unaccepted answers contains more value than the accepted one, while it may be not that meets questioner's requirements.
Thus, the scores of answers are used as our indicator for answer quality, where the authors of high quality posts can be considered as expert to specific query. Accordingly, here we keep only top-$k$ voted answers with a threshold $v$, which make sure selected answers, which essentially is our knowledge base, contains only answers scored more than $v$. 
Here, our system is actually rely on the voting system of Stack Overflow. We assume that the system can deliver accurate evaluation to answers.
As mentioned before, any entry-class qualified users can vote a post up or down.
Instance may occur that a non-professional user give negligent votes to posts. It is still elaborate to state our assumption, if taken account the large population of users.  

Still, the user-vote interactions are quite sparse data, which may not be perfect for our practice. Here supplements can really help, and we propound the employment of matrix factorization. 

Matrix factorization is a latent factors model, which to some extend can help with sparse data, which is widely used in industry, and adopted by many collaborative filtering recommendation systems (\cite{koren2009matrix,liang2016factorization,yao2015service,yao2018mashup}). It is also worth mentioning that a similar factorization technique, Tensor Decomposition, is also quite successful in these kind of applications(\cite{yao2018collaborative, huang2018expert}). However, it would be too computational expensive to update in our case here.
As the supplements to our context-based evaluation of user expertise, such latent information can further boost our accuracy.
Employing MF usually starts with a relatively sparse user-item matrix as input, and it de-composite the matrix into the user-latent factors multiplying with item-latent ones. In this work, it is used to learn latent voting information. 

Given matrix $V_{N\times M} = WH$, where $V$ is the answer-score matrix that contains voting information from $M$ users in $N$ questions(how many votes a user been given by posting answers to the question),
we apply the Non-negative Matrix Factorization(NMF) technique and define the loss function as:

\begin{equation}
  \pounds_{loss} = \argmin_{W,H}\frac{1}{2}\Vert X-WH\Vert^2_F = \frac{1}{2}\sum_{i,j}{(X_{i,j}-WH_{i,j})^2}
\end{equation}
where $\Vert\cdot\Vert^2_F$ is the Frobenius norm of the matrix.

Elastic Net regulator
combining $\ell$-1 and $\ell$-2 norms, along with parameter $\rho$ controls $\ell$-1 ratio and $\alpha$ regulates $\ell$-2 intensity, we have this regulation function:

\begin{align}
  \pounds_{reg} &= \alpha\rho\Vert W\Vert_1 + \alpha\rho\Vert H\Vert_1 \\       
                &+ \frac{\alpha(1-\rho)}{2}\Vert W\Vert^2_F + \frac{\alpha(1-\rho)}{2}\Vert H\Vert^2_F
\end{align}

Now, the objective becomes:

\begin{equation}
  J = \pounds_{reg} + \pounds_{loss}
\end{equation}

Making allowances for the completion of MF, we can apply the learned latent voting information.
In most case, voting shall occurs only within certain domains for one user, the latent voting data shall still be sparse. In practical, a weight $\lambda$ is introduce to regulate the combination of the learned latent information to the orignal data. After comparison the query $q$ with in domain $T$, top-$k$ answerers are finally output as experts.

\subsection{Time Complexity}
As indicated in Fig~\ref{fig:framework}, our approach contains two parts of offline processes, that is, the ``Post Representation'' and ``Expert Domain Extraction'', as well as one part of online process, that is, the ``Expert Recommendation". In application scenarios, where occasional updates may be necessary, once the initial offline preparation finished, these two offline processes can be done in the background with no influence to the running system. 
Similar structure has been applied before in other area of interests such as image retrieval and search, producing satisfying results(\cite{zhu2014effective, zhu2015topic, zhu2016learning, xie2016unsupervised}).
Hence, our discussion to time complexity in this work is about the online process, ``Expert Recommendation" part of our framework. 

Arguably, the is part of processing can be further partitioned into 3 subprocesses, which is: 1) new question vectorization, 2) new question domain extraction, and 3) inter-domain candidate matching.Since the word vectors are pre-trained, for 1) we need only a traversal of the new question, to find and summarize word vectors accordingly, and for a new question of length $l_q$, where the word vectors in our case is stored in a hashed data structure, the process can be done in $\mathcal{O}(l_q\times 1 ) ~ \mathcal{O}(l_q)$; Similarly, note the number of domains we have as $n_d$, an iteration can solve the domain extraction based on our offline prepared data, and this end up with $\mathcal{O}(n_d)$; for a domain contains $m_d$ existing sufficiently answered questions whose average number of high quality answers are $n_a$ , matching process in 3) can be a sequential iteration of $m_d$ and $n_a$, which results in an $\mathcal{O}(m_d+n_a)$. Thus, the time complexity of our approach, in online stage, is  $\mathcal{O}(l_q + n_d + m_d + n_a)$.

Considering in real cases and our dataset, the length of questions($l_q$) can barely excess a few hundreds of words, and the number of domains($n_d$) as well as average number of high quality answers to each existing questions($n_a$) are often numerically limited(see Section~\ref{subsec:dataset} below) we can safely simplify the who time complexity down to $\mathcal{O}(m_d)$ for approximation.

\section{Experiments}
\label{sec:experiment}
\subsection{Dataset}
\label{subsec:dataset}

\begin{table}[ht!]
  \centering
  \caption{Data Description for 3-fold tests}
  \label{tab:stat}
  \begin{tabular}{ll}
  \toprule
  Statistics                 & Value  \\ \midrule
  number of questions        & 118321 \\
  number of users            & 99220  \\
  number of answers          & 428370 \\ \bottomrule                
  \end{tabular}
 \end{table}

Training on extracted text corpus consists of 3,700,968,585 words from post title, body, and comments, a set of word representations with a vocabulary of size 1,346,955. Unpreventably, phrase like ``ping-test'' or user names have not been removed by preprocessing and can still be found in the vocabulary. Despite the impact of those words on trained representations can be ignored, such irrelevant words can still waste computing resource on post represenation generation and expert domain findings. This issue can be addressed by applying filters. Since Stack Overflow is a software programming websites, database from Tian's work (\cite{tian2014sewordsim}) to remove the non-software-related words is expedient. After filtering, vocabulary size dramatically dropped to 5,336.
Also, as mentioned, we kept only their top-5 voted answers of 11,832 satisfied questions(of the totally 5,916,073 data source questions). Moreover, a set of 3-fold tests are conducted using randomly chosen 100, 200, 300 and 400 queries, which are not in the selected questions, where all users involved are contained.

\subsection{Results Analysis and Evaluation}

\begin{table*}[!ht]
  \centering
  \begin{threeparttable}
  \caption{Accuracy comparison at top-5, with STM, SSRM, BPMF, PMF and Jaccard}
  \begin{tabular}{llllllll}
  \toprule
        & & Jacccard & PMF    & BPMF   & SSRM   & STM & \textbf{Ours}\tnote{1}\\\midrule
  accuracy@1 & & 0.0158   & 0.0045 & 0.0056 & 0.0578 & \textbf{0.1034} & 0.0581          \\
  accuracy@2 & & 0.0254   & 0.0045 & 0.0056 & 0.0765 & \textbf{0.1051} & 0.0914          \\
  accuracy@3 & & 0.0315   & 0.0067 & 0.0067 & 0.0810 & \textbf{0.1192} & 0.1021          \\
  accuracy@4 & & 0.0351   & 0.0078 & 0.0100 & 0.0836 & 0.1200          & \textbf{0.1283} \\
  accuracy@5 & & 0.0399   & 0.0089 & 0.144  & 0.0856 & 0.1267          & \textbf{0.1367}\\ \bottomrule
  \end{tabular}
  \begin{tablenotes}
      \item[1]{our approach is set with $\lambda = 0.5$ and the results are tested with 3-fold queries at size of 200}    
  \end{tablenotes}
  \end{threeparttable}
  \label{tab:prec}
  \end{table*}

\begin{figure}[!ht]
  \centering
  \includegraphics[width=0.8\linewidth]{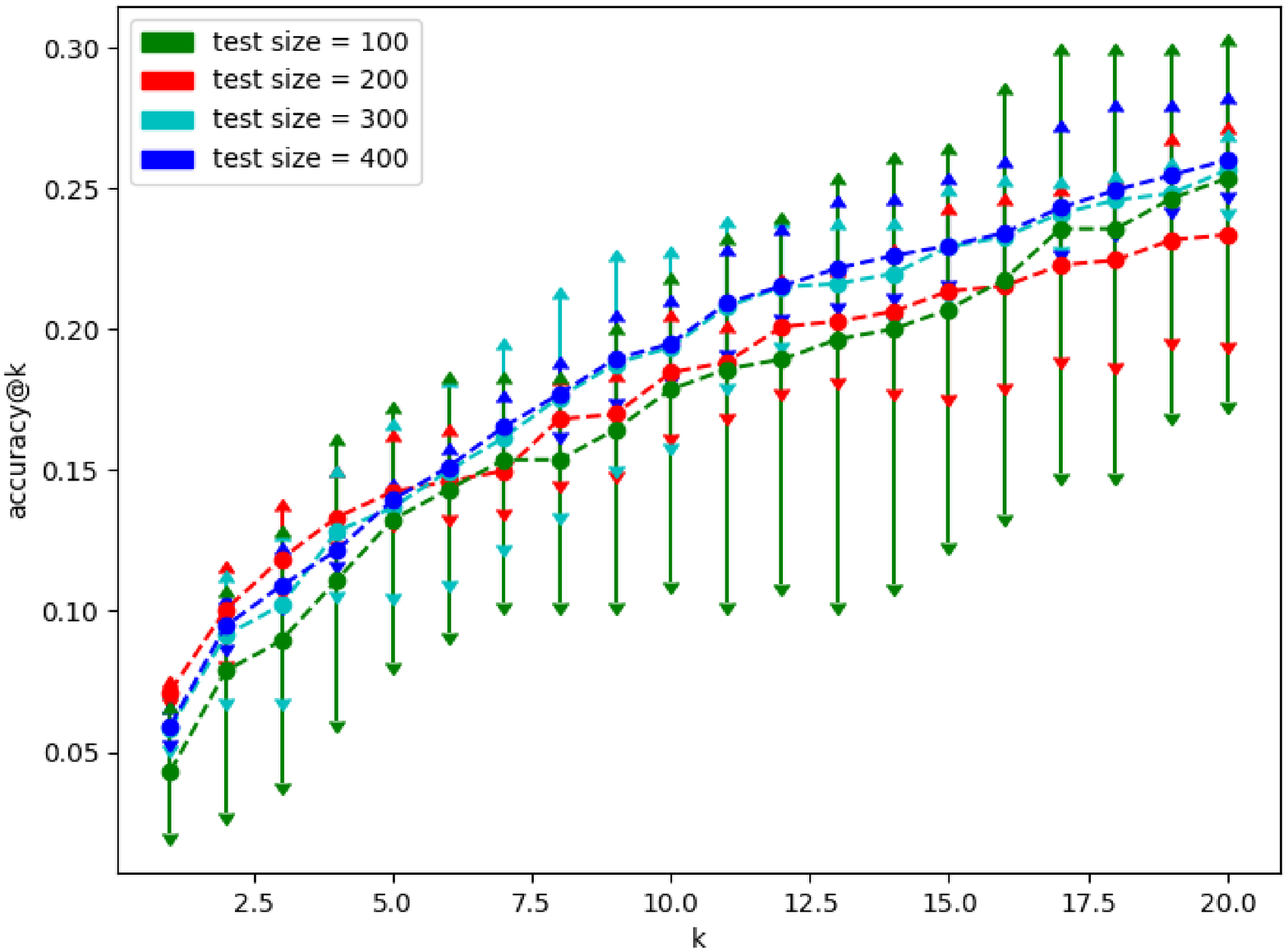}
  \includegraphics[width=0.8\linewidth]{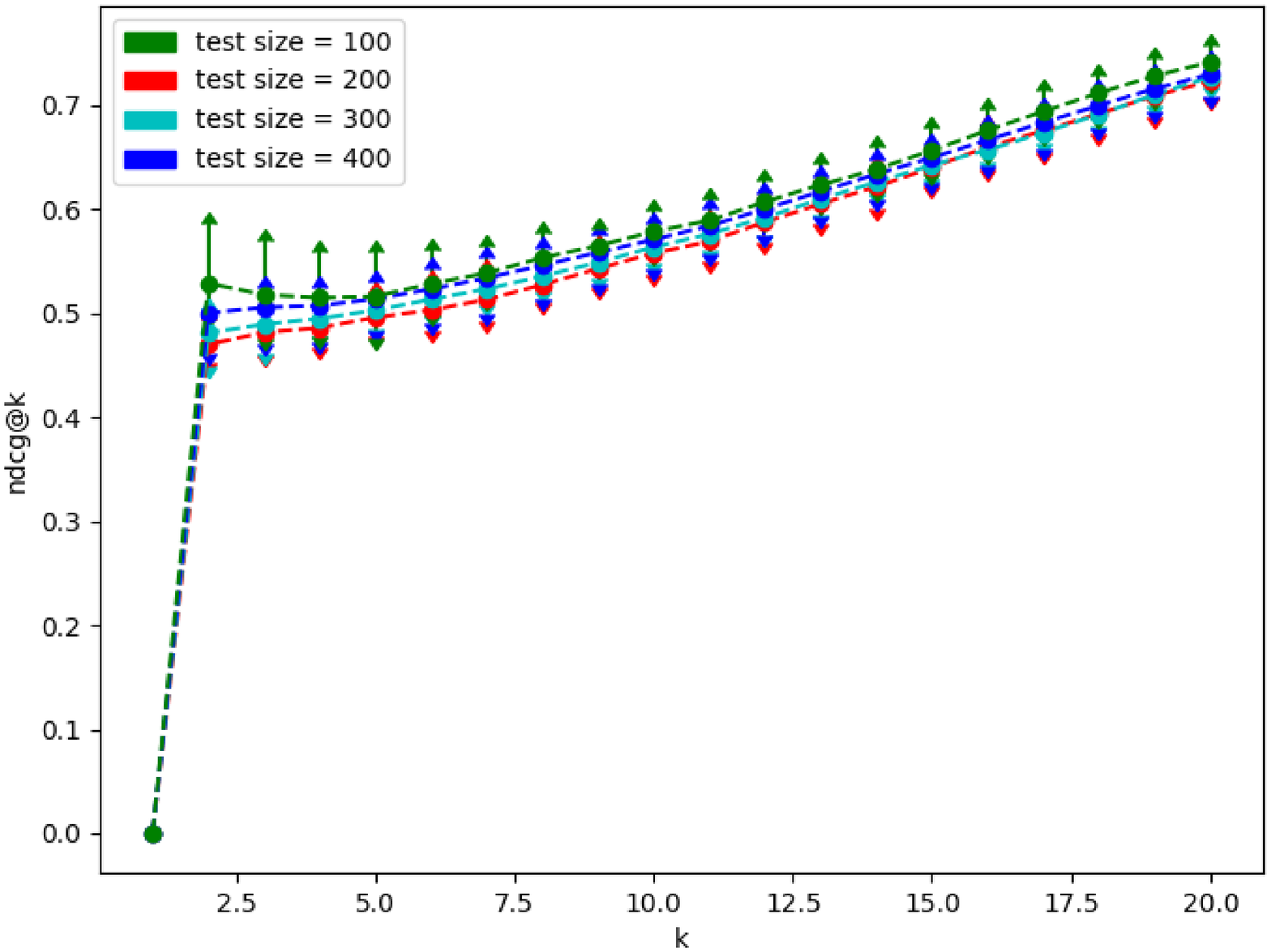}
  \caption{Accuracy and NDCG at top-k of 3-fold tests, with different test size, $\lambda = 0.5$}
  \label{fig:result}
  \end{figure}

It is obviously that the number of clusters can influence the accuracy of recommendation, for which Silhouettes Score are chosen as the measurement to evaluate the clustering process and help determine the optimal number of clusters. Silhouettes Score considers both density of a cluster and the separation between clusters. Also, we tried different $\lambda$'s, the weight we use to combine our expertise score with the supplemental information learned from matrix factorization, with a 3-fold test.

\begin{figure}[!ht]
    \centering
  \includegraphics[width=0.85\linewidth]{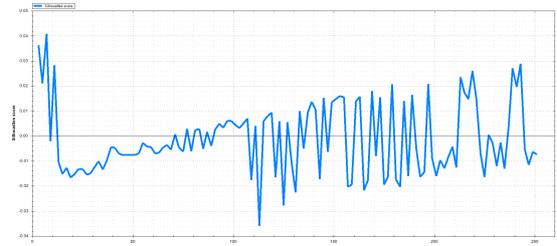}
  \caption{Silhouettes score for different cluster number \ref{fig:silscore}}
  \label{fig:silscore}
  \end{figure}

\begin{figure}[!ht]
    \centering
  \includegraphics[width=0.85\linewidth]{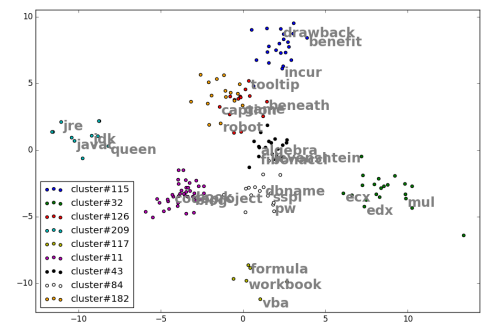}
  \caption{Example of selected word clusters, all points with same colour shown above belong to one cluster}
  \label{fig:eg_word_cluster}
  \end{figure}

Here, k-means algorithm are used for clustering and Silhouettes Scores are computed to difference numbers of clusters. Figure \ref{fig:silscore} shows the score at certain clustering numbers and Figure \ref{fig:eg_word_cluster} shows examples of domains with selected words. Noticed that very small number of clusters can produce very high Silhouettes score, yet, if such small number of clusters is optimal, it would be pointless to cluster at first place. Thus a cluster number 243 results in a acceptable Silhouettes score at 0.028879744, and this was latter proven optimal in our case.
As for the whole system, we use the precision@N(see~\cite{guo2008tapping} for more details) to measure accuracy and nDCG to assess recommendation quality.
The baseline here we compared is Jaccard similarity based approaches with a procedure of similar idea to our framework apart from word representation part.  Probabilistic Matrix Factorization(or PMF) and Bayesian PMF (or BPMF) (\cite{salakhutdinov2007probabilistic})is also tested with similar experimental setup, where both are enhanced version to basic matrix factorization approach.
We additionally compared two state-of-the-art methods, that is STM by \cite{Riahi:2012:FEU:2187980.2188202} and SSRM by \cite{dong2015predicting}.

Based on our experiments, it is believed the proposed framework of expert recommendation out-performs baselines and state-of-the arts approaches(see Table~\ref{tab:prec}). Matrix Factorization based techniques in these experiments end up not very effective, likely due to our super-sparse dataset.

Stabilty of proposed method is also tested in this work. Experiments are conducted with different test sizes(100, 200, 300 and 400 queries). Figure~\ref{fig:result} indicates our stable performance both in accuracy and quality for recommending top-20 experts.

\section{Conclusion}
\label{sec:conclusion}
In this paper we have proposed a framework to recommend potential experts, who may solve question in Q\&A website, or be the candidate of business recruitments.
Embedding techniques is utilized to generate representations and knowledge domains are extracted.
New query is also directed to go through the same process and mapped into the same linear space to compare, and expert behind posts are ranked and listed for recommendation.
Comprehensive tests are conducted and demonstrated our stable merit performance over certain existing approaches. 

\section*{Acknowledgments}
This research was undertaken with the assistance of resources and services from the \textit{UNSW Leonardi Engineering Research Cluster}(decommissioned) and \textit{National Computational Infrastructure} (NCI), where the latter is supported by the Australian Government.

\bibliographystyle{model2-names}

\end{document}